\documentclass[sigconf]{acmart}

\setcopyright{acmlicensed}

\renewcommand{\paragraph}[1]{\medskip\noindent\textbf{#1:~}}
\usepackage{array}

\newcolumntype{M}[1]{>{\centering\arraybackslash}m{#1}}
\newcolumntype{N}{@{}m{0pt}@{}}
\usepackage[hide]{chato-notes}
\usepackage{multirow}
\usepackage{balance}
\usepackage{siunitx}

\usepackage{tabularx}
\usepackage{makecell}
\usepackage{arydshln}
\usepackage{amsmath}
\usepackage[normalem]{ulem}
\usepackage{algorithm,algpseudocode}
\usepackage{graphicx}
\usepackage[para]{footmisc}

\acmConference[WWW '21]{WWW '21: International World Wide Web Conference}{March 8-12, 2021}{Ljubljana, Slovenia}

\usepackage{caption}
\usepackage[labelformat=simple]{subcaption}

\newcommand{\wang}[1]{\textcolor{black}{#1}}
\newcommand{\xwang}[1]{\textcolor{black}{#1}}
\newcommand{\xw}[1]{\textcolor{black}{#1}}

\newcommand{\xiwang}[1]{\textcolor{black}{#1}}

\newcommand{\yaya}[1]{\textcolor{black}{#1}}
\newcommand{\iadh}[1]{\textcolor{black}{#1}}
\newcommand{\craig}[1]{\textcolor{black}{#1}}

\newcommand{\xiw}[1]{\textcolor{black}{#1}}
\newcommand{\xiwa}[1]{\textcolor{black}{#1}}
\newcommand{\craigw}[1]{\textcolor{black}{#1}}

\usepackage{marginnote}

\begin{document}
%
\title{Leveraging Review Properties for Effective Recommendation}
\author{Xi Wang}
\affiliation{%
  \institution{University of Glasgow, UK}
}\email{x.wang.6@research.gla.ac.uk}

\author{Iadh Ounis}
\affiliation{%
  \institution{University of Glasgow, UK}
}\email{iadh.ounis@glasgow.ac.uk}

\author{Craig Macdonald}
\affiliation{%
  \institution{University of Glasgow, UK}
} \email{craig.macdonald@glasgow.ac.uk}

\begin{abstract}
\looseness -1 Many state-of-the-art recommendation systems \iadh{leverage} explicit \iadh{item} reviews \iadh{posted by users} by considering their usefulness in representing \iadh{the users'} preferences and describing \iadh{the items'} \xwang{attributes}. 
\iadh{These posted} reviews \craig{may} have \xwang{various} associated properties, such as their length, their age since \iadh{they} were posted, \iadh{or} their item rating. However, \iadh{it remains unclear how these different review properties contribute to the usefulness of their corresponding reviews in} addressing the recommendation task. In particular, 
users show distinct preferences \iadh{when considering different aspects of the reviews (i.e. properties)} \iadh{for} making decisions 
\iadh{about the} items. 
\iadh{Hence}, it is important to model the \iadh{relationship} between \iadh{the} reviews' properties and the usefulness of reviews \iadh{while} 
learning the users' preferences and the items' \xwang{attributes}.  \craig{Therefore}, we propose to 
model \iadh{the} reviews with \iadh{their} associated available properties. 
We \iadh{introduce} \iadh{a novel} review \iadh{properties-based} recommendation model (RPRM) that learns which review properties are more important than others \iadh{in capturing the usefulness of reviews}, thereby enhancing the recommendation results. 
\craig{Furthermore, \iadh{inspired by the users' information adoption framework}, we integrate two loss \iadh{functions} and a negative sampling strategy into our proposed RPRM model, to \iadh{ensure that the properties of reviews \xwang{are}
correlated with the users' preferences}.} 
We examine the effectiveness of RPRM using \iadh{the \craig{well-known}} Yelp and Amazon datasets.
\wang{Our results show that RPRM \iadh{significantly} outperforms a classical and five state-of-the-art \iadh{baselines}. Moreover, we experimentally show the advantages of using our proposed loss functions and negative sampling strategy, which further enhance the recommendation performances of RPRM.}


\end{abstract}

\maketitle

\section{Introduction}\label{sec:introduction}
\iadh{In recent years, there has been an increase} in the amount of available information and interaction choices online. As a consequence, recommender systems are increasingly being deployed in various platforms to alleviate the complexity of decision making for users and help them to find their \iadh{desired} items. \iadh{Several} studies~\cite{zheng2017joint,chen2018neural} \iadh{focused on leveraging the} item reviews posted by users. For example, 
Li et al.~\cite{DBLP:conf/ijcai/LiNLCQ19} \iadh{modelled the} users' dynamic preferences by first aggregating the reviews of users \iadh{and the reviews of the items they interacted with} in a time-sequential manner. \iadh{They converted} these reviews into embedding vectors to represent \iadh{the} users' preferences. However, \wang{not all reviews can be useful to represent \iadh{the} users' preferences and items' \xwang{attributes}~\cite{chen2018neural}. \wang{\iadh{Instead, by estimating the usefulness of reviews}, the recommendation models can \iadh{focus on those} valuable reviews \iadh{among the large volume of available information,}} \iadh{thereby} \iadh{leading to improved recommendation performance}~\cite{guan2019attentive,bauman2017aspect}. Moreover,} the performances of review-based recommendation models can be limited if they capture the users' preferences and items' \xwang{attributes} by solely using the review text~\cite{DBLP:conf/sigir/SachdevaM20}. \iadh{Hence, \wang{many studies \iadh{aimed} to incorporate the usefulness of reviews in review-based recommendation~\cite{chen2018neural,bauman2017aspect}}}. \craig{Some other approaches use an} attention mechanism to model the usefulness of reviews~\cite{guan2019attentive} or \iadh{those portions} of the textual content of reviews \iadh{that} contribute \iadh{most} to the recommendation performances~\cite{chen2019personalized}. 
However, \wang{we argue that} a limitation of \craig{such approaches} is that they capture the usefulness of reviews by relying on historical data, which \iadh{often do not generalise} to \wang{\craig{reviews that are unseen} by the trained model}~\cite{GraphAttnNetwork}. 

\wang{To address the limitation above, we propose to consider {\em review properties} to model the usefulness of reviews.} \xwang{\iadh{Indeed, the reviews posted by users on items have} a \iadh{corresponding set of properties, such as their length,  the number of days since they were posted (i.e.\ age) or \craig{their} writing style}. 
\xwang{These review properties} are associated with \iadh{the} historical reviews \iadh{as well as the} unseen reviews \iadh{by} the trained model. \iadh{A number of} studies~\cite{raghavan2012review,wang2019comparison} have \iadh{previously attempted to} leverage \iadh{the} review properties \iadh{when making} recommendations.}
\iadh{The underlying premise of such studies is that the 
review properties} \iadh{encapsulate} rich information \iadh{about both} the users' preferences and the items' \xwang{attributes}. 

\iadh{In particular, each review property can bring useful insights about the users' preferences and the items' attributes.} Therefore, by integrating \craig{such} review properties into the review modelling process, a review-based recommendation model could \iadh{also encapsulate}
the usefulness of reviews \iadh{when capturing the users' preferences} and \xwang{items' attributes}. \iadh{In the literature}, \iadh{a number of review properties have been used} as side/contextual information to enrich the user-item interactions \craig{when addressing} recommendation \craig{tasks}. For instance, the geographical \craig{property} of reviews \craig{have} been used to capture \iadh{those venues visited by a user} or \iadh{to} estimate \iadh{the} users' locations \iadh{when making} local recommendations~\cite{liu2019geo,manotumruksa2018contextual}. The \xiw{
temporal} property of reviews has been frequently \iadh{leveraged in sequential recommendation to} predict the next action of users~\cite{zhao2019go,wu2020deja,manotumruksa2018contextual}. \wang{However, these studies do not consider \iadh{the} review properties to examine the usefulness of reviews so as to improve the recommendation performances.}
\iadh{In particular}, it remains unclear how these different review properties contribute to \iadh{estimating the usefulness} of their corresponding reviews in \iadh{effectively} addressing the recommendation task. \xiw{We address this limitation by considering various review properties and \yaya{examining} their \yaya{actual} effectiveness in capturing the \yaya{reviews' usefulness in addressing} the recommendation task.}
\xiw{In~\cite{sussman2003informational}, Sussman et al. proposed the users' adoption of information framework, which \craig{showed} that} 
each user \iadh{follows} a particular scheme or strategy \iadh{in using} different \iadh{properties or aspects of the review} information and thereby \iadh{each user makes} different interaction decisions. For example, according to the Elaboration Likelihood Model (ELM)~\cite{filieri2014wom}, users \craig{can} follow two strategies to process \iadh{the} posted reviews, \iadh{namely} the {\em central route} and the {\em peripheral route}. Users that follow the central route \iadh{show} stronger \iadh{willingness in processing} in-depth information (e.g.\ the descriptions of items in \iadh{the} reviews), while \iadh{users who adopt the peripheral route} frequently use the overall ranking or rating as key factors to make decisions. \xiw{\yaya{The
various reviews posted by the users differ in their characteristics (e.g. length, language, details). Such differences can be \craigw{described} by the reviews' properties}, \yaya{and are} correlated to the level of \craigw{the} users' adoption of information~\cite{chan2011conceptualising,wu1987susceptibility,qahri2019factors}.} Therefore, \iadh{we argue that} a model can \iadh{better capture the users' preferences by learning} how users use the \yaya{reviews} and \iadh{by examining their preferences} on different properties of the reviews. 


\iadh{In this paper, we \xwang{model}} the \wang{
\xwang{importance} of different review properties in capturing the usefulness of reviews \iadh{and learning}} \iadh{the} users' preferences and the items' \xwang{attributes}. 
\looseness -1 We propose a novel review property-based neural network model (RPRM) to \iadh{effectively} address the recommendation task. RPRM investigates the usage of review properties to model the usefulness of reviews and aims \iadh{to enhance} the ability of \iadh{the} recommendation model in capturing the usefulness of reviews and \craigw{the} \xiw{users' adoption of information.} In particular, RPRM uses six \iadh{commonly encountered} review properties \craig{in recommendation scenarios} 
to encode the usefulness of reviews, including the age \iadh{of the} reviews, the length of the reviews, the \iadh{reviews'} associated ratings, \xw{the number of helpful} \iadh{votes associated to the} reviews, the probability of \iadh{the} reviews being helpful and the sentiment expressed in the reviews.  
\xiw{Note that in this paper,  `helpfulness' \yaya{refers to whether users found the reviews helpful for their task, for example by capturing if they voted them to be helpful}.} 
\wang{Different review properties can have various \xwang{importance} \iadh{levels} 
in capturing the usefulness of reviews for different users and items. Moreover, according to the \iadh{aforementioned} users' adoption of information \xwang{framework} ~\cite{sussman2003informational}, the properties of reviews are correlated \xiw{to} the level of users' adoption of information. This \iadh{suggests} that users tend to prefer items whose associated useful reviews capture the same important properties as those the users prefer. Therefore, we also propose two loss functions and a negative sampling strategy that aim to reward the \iadh{situation where} a user and the interacted items \iadh{agree} on the most important properties and penalises the \iadh{situation} where the user disagrees with the negative sampled items on the most important properties.}


\looseness -1 The main contributions\footnote{\xiwa{We will release all our code upon the acceptance of this paper.}} of this paper are: 

\textbf{(1)} \wang{We propose a novel review-based recommendation model, RPRM, which \iadh{leverages} the usefulness of reviews to address the recommendation task. To the best of our knowledge, this is the first \iadh{work} that integrates \yaya{the} review properties \xwang{in estimating the}  \yaya{reviews'} usefulness in a recommendation model \xiw{\yaya{through leveraging how users make use of such reviews in their interaction with the system.}}} 
\textbf{(2)} \wang{\iadh{Inspired by the users' adoption of information framework}, \xiw{we} propose two loss functions and one negative sampling strategy that model the agreement on the importance of review properties between \iadh{the} users and items. }
\textbf{(3)} \wang{We show that RPRM significantly 
outperforms one \iadh{classical} and five state-of-the-art recommendation approaches \craig{on the commonly-used Amazon and Yelp datasets.}}
\textbf{(4)} \wang{\iadh{We show that our} proposed loss functions and the negative sampling strategy can further enhance the recommendation performances of RPRM \iadh{across the two used} datasets. } 

\section{Related Work}
\looseness -1 \iadh{We briefly discuss three bodies of related work, namely recommendation approaches based on reviews, recommendation approaches leveraging the use of review properties, and work investigating users' behaviour while interacting with information}.

\subsection{Review-based \iadh{Recommendations}}
The main objective of applying a recommendation model is to observe \iadh{the} users' behaviours and to learn \iadh{how to distinguish} among items \iadh{for a given user}, thereby estimating \iadh{the users'} preferences and recommending \iadh{suitable} items \craig{that} the \iadh{users might} be interested in. User-generated reviews \iadh{encapsulate} rich semantic information \iadh{such as the possible explanation of the users' preferences and the description of specific} item \xwang{attributes}~\cite{chen2015recommender}. Therefore, many recommendation models \craig{have} \iadh{aimed to leverage these} reviews to construct user/item representations and to address the 
recommendation task\xiw{~\cite{chen2020local,he2015trirank,almahairi2015learning,DBLP:conf/ijcai/LiNLCQ19,ni2019justifying}}. \iadh{Many} \iadh{previous} review-based recommendation \iadh{approaches} captured \xwang{the semantic similarity} 
between \iadh{the \xiw{review} content} ~\cite{zheng2017joint,chen2018neural}, which \iadh{allows to encode} additional relationships among \iadh{the} users and items, \iadh{\xiw{allowing} to better suggest items the users might be interested in}. \iadh{Indeed}, the posted reviews by users are valuable in modelling the \iadh{interactions} among users and items from \iadh{a} textual semantic perspective.
\wang{However, 
the quality and usefulness of \iadh{the} reviews \iadh{markedly vary with} the increasing amount of \iadh{users and the available reviews they post online}.}
Therefore, Chen et al.~\cite{chen2018neural} applied \iadh{an} attention mechanism to
\iadh{estimate} the usefulness of different reviews. \xwang{\iadh{Unlike previous work}~\cite{chen2018neural,bauman2017aspect} 
, which used \iadh{an} attention mechanism to \xiwang{learn} the usefulness of reviews, 
we argue that \iadh{the} review properties can be \iadh{directly leveraged} to effectively \xiwang{\iadh{capture} the usefulness of reviews}. 
} \wang{Moreover,} 
there are many \iadh{existing} approaches~\cite{raghavan2012review,manotumruksa2018contextual,wang2019comparison} that extract \iadh{the} review properties and \iadh{integrate them} as side or contextual information to enhance \iadh{the recommendation performance}. \iadh{However, unlike our work in this paper, such approaches do not make use of the reviews themselves}. \iadh{In the following, we further describe such approaches using the properties as side information.}

\subsection{\iadh{Recommendations using Review Properties}}\label{ssec:side_info}
\iadh{Various existing approaches\xiw{~\cite{ren2017social,ling2014ratings,raghavan2012review}} aimed to} leverage different review properties \iadh{as side information} to model \iadh{the} users' behaviours or \iadh{the items'} \xwang{attributes}. For example, Raghavan et al.~\cite{raghavan2012review} leveraged the \iadh{extent to which a review is helpful} to measure the reliability of the associated users' ratings and \iadh{to} incorporate such reliability scores \iadh{into} a recommendation model. 
The geographical property of \iadh{the} users' reviews~\cite{qian2019spatiotemporal,liu2019geo} \iadh{has} also been well studied in venue recommendation models. \iadh{Another} \iadh{property is whether a review expresses a sentiment}. 
For instance, Wang et al.~\cite{wang2019comparison} \iadh{replaced} the explicit ratings \iadh{provided by users} with the \iadh{review} sentiment scores to enhance the recommendation performance.  \iadh{The} temporal and age properties of reviews have been integrated into various recommendation models, and especially into the sequential recommendation models~\cite{zhao2019go,wu2020deja,manotumruksa2018contextual}. \iadh{For example,} Manotumruksa et al.~\cite{manotumruksa2018contextual} \iadh{encoded} the temporal information \iadh{in} a recurrent neural network to model \iadh{the} users' dynamic preferences \iadh{in the venue recommendation task}. 
\xiw{In particular, unlike previous works that have used review properties solely as side information \yaya{in a recommendation model}, in this paper we instead use them to estimate a given review's usefulness in enhancing \yaya{the performance of a recommendation model}.}
\xiw{Additionally, the distinct focus of various review properties is correlated to the users' adoption of information~\cite{chan2011conceptualising,wu1987susceptibility}.}

\looseness -1 \iadh{\xiw{Furthermore, although} \craig{these}  approaches} extracted various review properties as contextual information \iadh{in order} to \iadh{further} enrich the collaborative interactions among users and items, \wang{\iadh{they} \craig{have ignored} the textual information of reviews.}  \iadh{Indeed,}
\xwang{while} collaborative approaches can be effective with rich interaction information \xiwang{and side information}~\cite{manotumruksa2018contextual,zhang2015geosoca}, \wang{\xiwang{we argue that} it \iadh{is still} important to use both the textual information of reviews and their associated review properties. 
\craig{Indeed}}, \xwang{in this study,} we \iadh{postulate} that \iadh{leveraging the} review properties (e.g.\ length, age, sentiment) \iadh{and their relationships to the users' preferences can help} the \iadh{recommendation} model to \iadh{more accurately} learn the usefulness of reviews \iadh{for effective recommendation}. 
\iadh{\craig{In particular}, an effective recommendation} model \wang{needs to also} capture the users' preferences and the items' \xwang{attributes} along with the usage of \wang{these} \iadh{reviews' properties}. For instance, a user \iadh{might} prefer \iadh{to read} recent reviews \iadh{on a hotel} to \iadh{obtain a more accurate information on its} current condition \iadh{and services} instead of reading \iadh{much older} reviews. \iadh{This \iadh{intuition}} also aligns with the users' adoption of information \xwang{framework} \iadh{described in the next section}. 

\vspace{-2mm}
\subsection{Users' \iadh{Adoption of Information}}\label{ssec:user_adopt_info}
\iadh{Information adoption concerns} \craig{how} consumers modify their behaviour by \iadh{making use of} the suggestions made in online reviews~\cite{sussman2003informational}. The communication routes and \iadh{the customers'} involvement in a consumer opinion sharing website \iadh{might persuade} a customer to visit a particular destination or purchase a specific product~\cite{tang2012dual}. \iadh{A number of \xiwang{user}
\xwang{studies}} 
have examined various properties of reviews that influence the users' adoption of information~\cite{chan2011conceptualising,filieri2014wom,wu1987susceptibility,qahri2019factors,cheung2012review}. \wang{These} studies observed \iadh{that} the properties of reviews are correlated \xiw{to} the level of users' adoption of information -- indeed, such correlations are important motivations for our \iadh{present} work. 
\wang{For example,} Filieri and Mcleay~\cite{filieri2014wom} \iadh{used} the Elaboration Likelihood Model (ELM)~\cite{petty2012communication} to group the factors and properties of reviews according to two information processing routes (i.e.\ central and peripheral routes). 
\craig{The same authors also} observed that a peripheral route-based user would prefer to process the information \iadh{about} a product \iadh{that} simply has a good overall ranking. 
\looseness -1 \iadh{On the other hand}, a central route-based user would consider in-depth information to make decisions. 
Furthermore, Sussman et al.~\cite{sussman2003informational} considered the information usefulness as a mediator between the information process and the information adoption \iadh{by users} and \wang{showed} a strong linkage between the usefulness of the information and the \iadh{users'} decision making. \iadh{Our work is inspired by the aforementioned users' adoption of information \xwang{framework}}.
\iadh{In particular, we argue that leveraging the users' preferences in relation to the reviews' properties can improve recommendation effectiveness by providing additional insights about the users' information processing behaviours and their personalised preferences on the item's attributes as conveyed by the reviews' properties. \xiw{\yaya{To the best of our knowledge, this is the first work that uses and leverages} the users' adoption of information to address the recommendation task.}}



\section{Methodology}
\looseness -1 We first introduce the recommendation task and the notations used in this paper. Next, we describe our proposed RPRM model, which \iadh{leverages the reviews posted by users to enhance the recommendation task}. \iadh{RPRM takes into account the} review properties \iadh{when} modelling \iadh{the} user/item information by learning the \iadh{importance of} different review properties \iadh{for enriching} the representations of the users' preferences and the items' \xwang{attributes}. \wang{\iadh{RPRM accounts for the importance} of \iadh{the} review properties for users and items by proposing two loss functions and a negative sampling strategy that} \xwang{model \iadh{the extent to which the users and items agree} on the \iadh{important} review properties. \iadh{For example, the agreement will be high if a user considers longer reviews to be more useful and a given item's reviews usefulness is better described by long reviews}}. 



\subsection{Task Definition}
\looseness -1 \iadh{We address} the recommendation task, which aims to effectively identify and recommend items to users according to their preferences. The recommendation task \iadh{involves} \craig{connecting two key entity types, namely:} 
the set of users $ U = \{u_1, u_2, ..., u_{N}\}$ with size N and the set of items $I = \{i_1, i_2, ..., i_{M}\}$ with size M. To address the recommendation task, we aim to accurately estimate \iadh{the} users' preferences on items \iadh{so that we rank the items \craig{that} a given user might find the most interesting} in higher ranks. \iadh{To do so}, we \iadh{investigate the use of the reviews \craig{that} the users have posted on items, as well as} their associated properties. 
Each user $u$ or item $i$ has an associated set of reviews, e.g. posted by that user, $C_u$, or posted on that item, $C_i$. 

\looseness -1 \iadh{Furthermore,} the reviews of \iadh{a user or an item} can be described \iadh{using} $k$ review properties $\textbf{P} = \{P_{1},P_{2}, ..., P_{k}\}$. For example, to \iadh{capture} the \iadh{preferences of} user $u$ \iadh{for a given} property $P_{1}$, we \iadh{estimate} the corresponding review property \iadh{scores} for each review in the review set of user $u$ -- i.e. $P_{1,u} = \{p_{1,1}, p_{1,2}, ..., p_{1,|C_{u}|}\}$\xw{, where \iadh{$p_{1,t}$} is the property score of the \iadh{$t^{th}$} review of user $u$}. \iadh{For example, for the length (or age) property, the score will correspond to the length of the review (or its age resp.)}.  
\iadh{These scores could be computed for any property, provided that the property values are mapped into scalars in the range of \craig{[0..1]} using an adequate function. For example, the geographical property (`near' \craig{vs.\ `distant'}), the length property (`long' vs.\ `short'), or the age of reviews (`old' vs.\ `recent') can all be mapped into a scalar in the interval [0, 1].} \xw{In particular, the property scores also define the usefulness of reviews. For example, for \iadh{the} length property, a longer review \iadh{will have its length property} score closer to 1 than \iadh{other} shorter reviews, \iadh{and hence this would indicate} according to the length property scores, \iadh{that} a longer review is more useful.} \iadh{The computed} property scores enable the modelling of reviews from different aspects and examine the relationship between the review usefulness and the review properties. 




\begin{figure}[h]
    \centering
    \includegraphics[width=1.0\columnwidth]{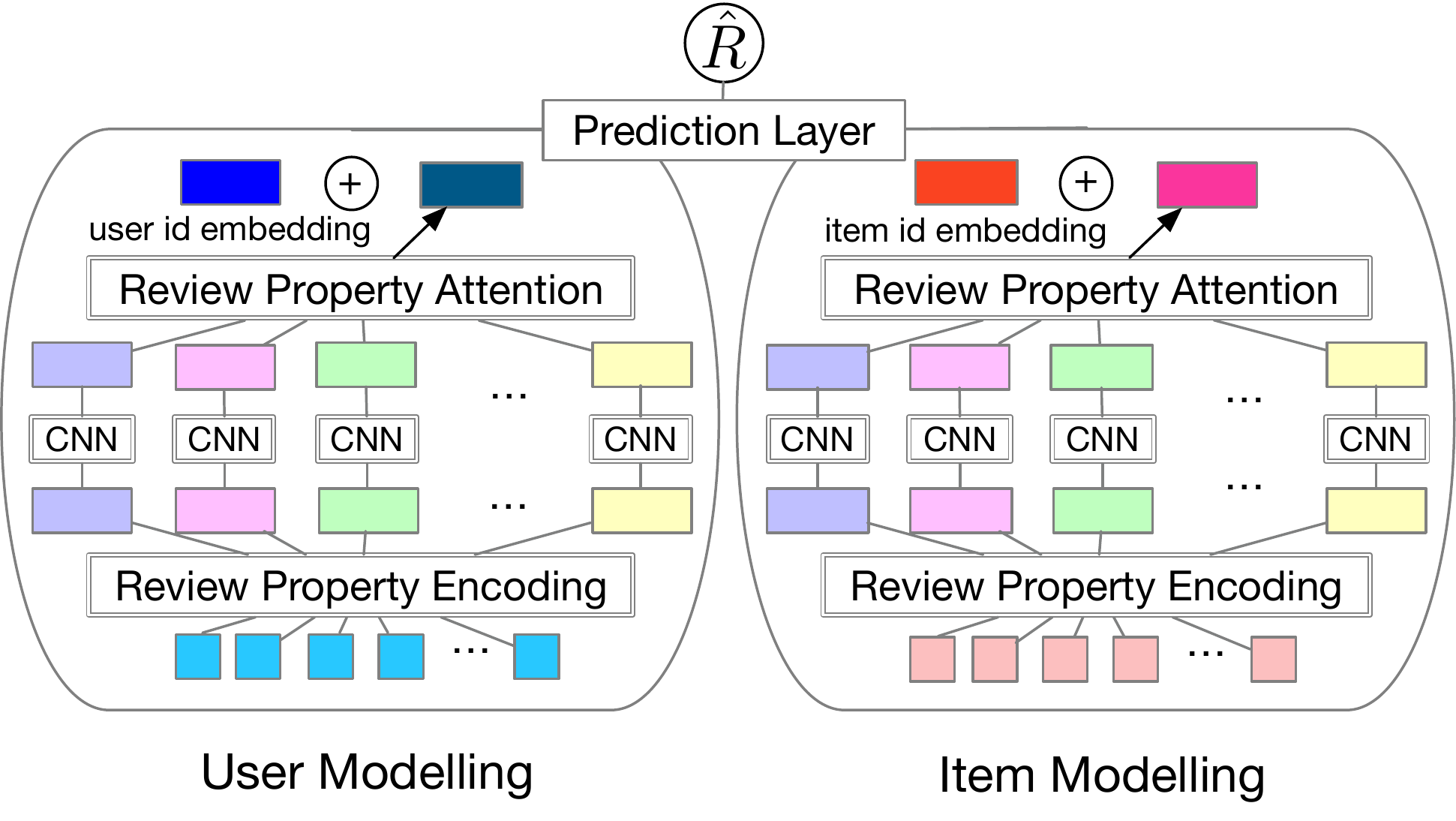}
    \caption{The Neural Network Architecture of RPRM.}\label{fig:RPRM}\vspace{-\baselineskip}
\end{figure}

\subsection{\iadh{The} RPRM Model}\label{ssec:RPRM_model}
To address the introduced recommendation task, we propose \craig{the novel Review Properties-based Recommendation Model (RPRM), which is a neural recommendation model that} takes reviews and their associated properties into account. In particular, we \iadh{use a} \xwang{dot-product attention} mechanism 
to \iadh{score} and learn which review property is more useful \iadh{in describing} the usefulness of reviews and thereby \iadh{are important in} making good recommendations. We first present the architecture of our proposed RPRM model in Figure~\ref{fig:RPRM}. In general, RPRM is a collaborative filtering-based framework, which \iadh{models} the interactions between users and items. \iadh{It is of note} that RPRM models \iadh{both the users and items using} the same neural network architecture (i.e.\ User and Item Modelling in Figure~\ref{fig:RPRM}). The \iadh{RPRM} architecture \iadh{is organised} into \craig{four layers}, which we discuss in turn below: (1) \iadh{The} review property encoding layer, \craig{which combines \iadh{the} semantic textual representations of reviews obtained using BERT with the properties of reviews} (Section 3.2.1); (2) \iadh{The} review embedding processing layer, \craig{which creates a low-dimensional representation of each review (Section 3.2.2)}; 
(3) \iadh{The} review property \xwang{attention} layer (Section 3.2.3), \craig{which identifies the properties of reviews \iadh{that} are more useful to represent the users' preferences and items' attributes}; (4) We use the output from the last layer as input to the prediction layer, along with the identification embedding of \iadh{a given user and item}, to score the user's \iadh{preferences} on items (Section 3.2.4). \craig{Later, in Section~\ref{ssec:model:learning}, we discuss how we propose new loss functions and a new negative sampling strategy to aid learning while encapsulating \iadh{the} properties of reviews.}

\subsubsection{Review Property Encoding Layer}


\looseness -1 RPRM first models the users' reviews and items' reviews. To process and summarise the semantic information of each review, we convert 
\xw{each reviews into a 768-sized embedding vector} by using the pre-trained BERT model~\cite{DBLP:conf/naacl/DevlinCLT19}, which is a recent \iadh{widely used} language modelling approach\iadh{\footnote{We use BERT \craig{for ease of} integration, but any language modelling approach could be used, e.g.\ \craig{ALBERT~\cite{lan2019albert} or RoBERTa~\cite{liu2019roberta}}.}}. 
\iadh{Next}, in this layer, the model encodes the embedding vectors of \iadh{the} reviews with various review properties \xwang{through a dot-product function}. The objective of encoding \iadh{the} review latent vectors with different review properties is to model the usefulness of reviews from different \iadh{perspectives (e.g.\ length, age, sentiment)}. 
Each review property can be \iadh{represented} by a list of normalised review property scores. \iadh{These} scores \iadh{allow} RPRM to focus on different reviews and encode the knowledge of the corresponding \iadh{reviews' properties}.
For example, by encoding the review length property, the model \craig{can} \iadh{capture} how reviews with different lengths can \iadh{have an influence on} the recommendation outcome and how the length property of reviews \iadh{is} associated with \iadh{the} user/item representations. The encoding process of \iadh{a given} review property can be \iadh{described as follows}:
\begin{equation}\label{equ:review_property_encode}
    O_{u,P_{1}} = [X_{1}P_{1,1}, X_{2}P_{1,2}, ..., X_{|C_{u}|}P_{1,|C_{u}|}]
\end{equation}
where $X_{1,...,|C_{u}|}$ \xw{are the embedding vectors} of \iadh{the} reviews of user $u$ and $|C_{u}|$ is the size of his/her review set. In particular, Equation~(\ref{equ:review_property_encode}) \iadh{encodes} the review property $P_{1}$ for user $u$. After encoding $k$ review properties, for user $u$, we have $O_{u} = [O_{u,P_{1}}, ...,O_{u,P_{k}}]$. 

\iadh{In this work, we use six commonly available review properties to describe \iadh{the}} reviews from different perspectives:

    \noindent\textbf{$\bullet$} \textbf{Age: } We calculate \craig{the 
    number of days} $d$ \iadh{since} a review has been posted. \yaya{Then, we compute the Age score of a review to be:} $p = 1- d / max(D)$, \craig{where $max(D)$ is the age of the oldest review in the collection.} \xiw{\craigw{In this case}, a recent review is considered more useful than an \craigw{older} review.} 
    
    \noindent\textbf{$\bullet$}\textbf{Length:} The number of words that are included in a review.
    
    \noindent\textbf{$\bullet$}\textbf{\xwang{Rating}:} The rating associated with the review \craig{(1-5 stars)}.
    
    \noindent\textbf{$\bullet$}\textbf{Polar\_Senti: } \xiw{The Polar\_Senti \yaya{property}
    indicates the probability of a given review being polarised (\yaya{strongly} positive or negative).} We \iadh{use} a CNN classifier, \xiw{which identifies reviews as \yaya{being} positive or negative and \yaya{which} has been validated as a \yaya{strongly effective }classifier with >95\% classification accuracy in~\cite{wang2019comparison}}. \yaya{We} obtain the corresponding probabilities of the positive reviews being \iadh{actually} positive or the negative reviews being negative\footnote{\iadh{A review} is positive (negative if it has a rating $\geq$ 4 ($\leq 2$). The polarity of a 3-star review is predicted by the CNN classifier.}. 
    
    \noindent\textbf{$\bullet$}\textbf{Helpful: } The number of helpful votes given by other \craig{users} to a \craig{particular} review.
    
    \noindent\textbf{$\bullet$}\textbf{Prob\_Helpful: } We classify \iadh{the} reviews with a state-of-the-art review helpfulness classification model~\cite{wang2020negative} and obtain the probability of \iadh{a given review to be} helpful.

In \iadh{addition}, for \iadh{the} Length, Rating and Helpful review properties, which have their property scores larger than 1, we apply \iadh{the} min-max normalisation to scale the property scores \iadh{into} [0, 1]. \xiw{\yaya{Note} that we use \yaya{the aforementioned} review properties as \yaya{typical review properties that are commonly available in various datasets}. However, \yaya{our approach is general in that it could also incorporate} other review properties (e.g.\ the geographical and part-of-speech properties) into \yaya{the} proposed RPRM model.}

\subsubsection{Review Processing Layer}
After encoding the embedding vectors of \iadh{the} reviews with \iadh{their} review property scores, we use the convolutional operators, as in other review-based deep neural network \iadh{approaches}~\cite{zheng2017joint,chen2018neural}, to model the embedding vector of each review. The convolutional operators consist of $m$ neurons, \craig{with} the $j^{th}$ neuron \craig{modelling} the review embedding vector \wang{as} $Z_{j} = ReLU(V * K_{j} + b_{j})$, where $V$ is the review embedding vector and $*$ is the convolution \xiw{operator} with the $j^{th}$ filter and \wang{$b_{j}$ is \craig{a} bias term}. \craig{The} ReLU activation function is applied to process the generated features. Next, \wang{each neuron $j$ applies} a sliding window over the features $Z$ with a max pooling function \iadh{to} then obtain the convolutional output $o_{j}$ for the corresponding neuron. Therefore, for each review, we concatenate the convolutional output from \iadh{the} neurons and obtain the processed embedding vector for each review (i.e.\ $O = [o_{1}, o_{2}, ..., o_{m}]$).


\subsubsection{Review Property \xwang{Attention} Layer}\label{sssec:attention_layer}
\looseness -1 In the review property encoding layer, RPRM converts \iadh{the} user/item modelling latent vectors into a set of latent vectors by considering various review properties. In this \xwang{attention} layer, the main objective is to observe which \wang{properties} of reviews \iadh{are} more useful to represent \iadh{the} users' preferences and items' \xwang{attributes}. \iadh{We hypothesise that} \xwang{the dot-product attention} mechanism \iadh{would} enhance the recommendation \iadh{performance} of RPRM. Moreover, each user or item is associated with a review property \xw{weighted} vector $\phi_{u}$ or $\phi_{i}$ with size $k$, \wang{where $k$ is the number of \iadh{used} review properties.} For a given user $u$, the review property \xwang{attention} layer \craig{is defined as}:

\begin{equation}
    O'_{u} = \frac{\sum_{t=0}^{k}\phi_{u,t}O_{u,P_{t}}}{k}
\end{equation}


\subsubsection{Prediction Layer}
\looseness - 1 In \iadh{this} layer, RPRM concatenates the processed review latent vectors and the identification embedding vector of users and items to make recommendations. The final prediction \iadh{of the users' preferences on items can be computed as:}
\begin{equation}
    \hat{R}_{u,i} = (O'_{u} \oplus V_{u}) \odot (O'_{i} \oplus V_{i}) 
\end{equation}
\looseness -1 where $\oplus$ is the concatenation operation, \iadh{which} combines the review embedding vector $O'$, and the identification embedding vector $V$. Moreover, $\odot$ denotes the element-wise product of the latent vectors between user $u$ and item $i$ to calculate the 
\wang{preference score $R_{u,i}$ of user $u$ on item $i$.}

\subsection{Model Learning}\label{ssec:model:learning}
\looseness -1 The RPRM model \iadh{addresses a} ranking-based recommendation task, \iadh{i.e.\ for a given user, it ranks first those items likely to be of interest to the user.} A common and popular \wang{ranking} scheme is to first apply the Bayesian Personalised Ranking (BPR) loss function~\cite{DBLP:conf/uai/RendleFGS09} to optimise the model by comparing the prediction scores \iadh{for} users $U$ with \iadh{the} positive items $I^{+}$ and \iadh{the} negative items $I^{-}$. The positive items are \iadh{those items the user has interacted with} \iadh{while the} negative items \iadh{are \xwang{sampled from} those items the users did not interact with thus} far. 
In particular, the uniform sampling strategy is \iadh{commonly} used to generate \iadh{the} negative items \wang{from \iadh{the} users' unseen items}. We \iadh{use} this learning scheme as a basic setup of our proposed RPRM model. 

 Aside from building upon \iadh{the} BPR's loss function and uniform sampling for generating negative items, we propose novel learning schemes to enhance the recommendation effectiveness. 
According to the
users' adoption of information \xwang{framework}~\cite{sussman2003informational} that we discussed in Section~\ref{ssec:user_adopt_info}, users \iadh{show} distinct information processing behaviour. 
\iadh{In the recommendation scenario,  users tend to have different preferences; for example some users might prefer shorter reviews while others might favour in-depth reviews that describe the advantages/disadvantages of a given item}. 
In particular, there is a \iadh{relationship} between \iadh{the users'} behaviour and \iadh{the} review properties~\cite{sussman2003informational}. 
\wang{This \iadh{suggests} that users tend to prefer items whose associated useful reviews capture the same important properties as those the users prefer. Therefore, we propose two loss functions (i.e. $PropLoss_{uu}$ and $PropLoss_{ui}$) that reward the case of a user and the interacted items agreeing on the most important properties and \iadh{penalise} the case where the user disagrees with the negative sampled items on the most important properties.}

In particular, \craig{based on} the users' adoption of information \xwang{framework}, we assume that users would \iadh{prefer} to process information from items \iadh{that have similar usefulness \xw{importance scores} 
on \iadh{the}} review properties. \wang{\iadh{Using the information from the similarly scored items' properties,}} \wang{\iadh{users} would exhibit \craig{a} higher probability of \craig{interacting} with these items than \iadh{with other} unknown items}. \wang{Therefore, both of our proposed loss functions ensure \iadh{that there is an} agreement in the importance of review properties between \iadh{the} users and their interacted items (i.e. positive items).}
However, $PropLoss_{uu}$ \wang{ amplifies} \wang{the disagreement in the importance of review properties} between \iadh{the} users and the \craig{unseen (negative) sampled} \iadh{items}, while $PropLoss_{ui}$ \iadh{amplifies} the disagreement
\wang{between the interacted items and the \craig{unseen (negative) sampled} items of users}. These two losses functions are defined \iadh{as follows}:
\begin{align}
    \begin{split}\label{equ:prop_loss_uu}
        &PropLoss_{uu}(u,i^{+}, i^{-}) = Sim(\phi_{u}, \phi_{i^{-}}) - Sim(\phi_{u}, \phi_{i^{+}}) 
    \end{split}
    \\
    \begin{split}\label{equ:prop_loss_ui}
        &PropLoss_{ui}(u,i^{+}, i^{-}) = Sim(\phi_{i^{+}}, \phi_{i^{-}}) - Sim(\phi_{u}, \phi_{i^{+}})
    \end{split}
\end{align}
where $Sim(.)$ is \craig{a} function that measures the similarity between the \iadh{weighted} vectors \wang{of \iadh{the} review properties}. \xwang{Before applying the similarity function, we scale the weighting scores by dividing the scores \iadh{by} the sum of scores in \iadh{each weighted} vector to generate} 
\xw{\iadh{a} \xiw{discrete probability} distribution \iadh{of} scores \iadh{[0..1]}} over \iadh{the} review properties. 
In particular, we \iadh{use} the \craig{Cosine} similarity (Cos) function and the Kullback–Leibler (KL) divergence \craigw{measure} as the similarity functions, which have shown good \iadh{performances} in measuring latent vector similarities~\cite{wang2019evaluating}. \xiw{Note that, since KL is a divergence measure, we use the inverse \yaya{of} KL to \yaya{compute similarity}.} 
Furthermore, we combine the PropLoss \wang{functions} with the commonly-used BPR loss \wang{function} as follows:
\begin{equation}
    \mathcal{L} = \alpha \times BPR(u,i^{+}, i^{-}) + (1-\alpha) \times PropLoss(u,i^{+}, i^{-})
\end{equation}
where $\alpha$ \iadh{controls} the \craig{emphasis} on \iadh{the} two loss functions.



\wang{\iadh{We} \iadh{also} propose a novel negative sampling strategy, called $Prop$ $Sample$, which models the agreement in the importance of review properties between \iadh{the users'} interacted items and the unseen items.} \wang{We argue that if \xwang{the same properties are important to} two items (e.g.\ \xwang{$i_1$ and $i_{2}$}), 
but \iadh{a particular} user interacts with item $i_1$ but not \iadh{with} item $i_2$, \iadh{then this} user \iadh{shows} a clearer preference \iadh{for} item $i_1$ \iadh{over} item $i_2$}. 
Therefore, we sample negative items from \craig{each user's} unseen items by selecting items \craig{that have similar review properties with those items the user has already interacted with.}

\looseness -1 For a given positive item $i^{+}$ $\in$ $I$, we \craig{again} use a similarity function $Sim()$ to calculate the similarity on the paired property weighted vectors $\phi_{i,p}$ between the positive item $i^{+}$ and all negative (unseen) items (i.e. $I^{-}$). Next, similar to the loss functions, we normalise the similarity scores across all negative items into a probability distribution. This probability distribution gives the likelihood for sampling these items as a negative instance for learning. 

\section{Experimental Setup}\label{sec:experimental_setup}
In this section, we examine the \iadh{performances of our} proposed \iadh{model and approaches} on two real-world datasets. Moreover, we compare the performance of RPRM with one \iadh{classical} and five state-of-the-art recommendation approaches. \iadh{In particular, we evaluate the performances} of our proposed loss functions and negative sampling strategy in addressing the following research questions:

\noindent\textbf{RQ 1:} Does RPRM \iadh{outperform} the recommendation \iadh{baselines} on \iadh{the two used} datasets?

\noindent\textbf{RQ 2:} \iadh{Do} the proposed loss functions, $PropLoss_{uu}$ and $PropLoss_{ui}$, improve the recommendation performances of RPRM \iadh{in comparison to the classical \wang{BPR loss} \iadh{function}}?

\noindent\textbf{RQ 3:} Does the proposed negative sampling strategy, namely $Prop$ $Sample$, \iadh{further} enhance the recommendation \iadh{performance} of RPRM \xwang{compared to \craig{the} uniform sampling strategy}? 

\subsection{Datasets \& Evaluation Metrics}\label{ssec:datasets}
\looseness -1\iadh{For answering} the aforementioned research questions, we use two real-world datasets, \iadh{namely} the Yelp dataset\footnote{https://www.yelp.com/dataset} and the Amazon Product dataset\footnote{http://jmcauley.ucsd.edu/data/amazon/}~\cite{mcauley2015inferring,he2016ups} to examine the effectiveness of our RPRM model \iadh{as well as} our proposed loss functions and negative sampling strategy. The Yelp dataset includes user reviews on \iadh{their} top popular category (i.e.\ `restaurant') and the Amazon dataset contains user reviews on products among six categories\footnote{`amazon instant video', `automotive', `grocery and gourmet food', `musical instruments', `office products' and `patio lawn and garden'}. \xiw{The use of various categories \yaya{of} the \craigw{Amazon} dataset allows to capture \yaya{the} users' preferences \yaya{across} different types of items/products.}
These two datasets have been used in \iadh{several} previous studies~(\iadh{ e.g. \cite{mcauley2013hidden,seo2017interpretable})}. \craig{We} \iadh{use} the Yelp dataset from the \iadh{most} recent round of Yelp challenge dataset (i.e.\ round 13). 



\looseness -1  In \iadh{our experiments}, we \xiwang{\craig{remove} cold-start users and items \craig{from} both datasets}, as in~\cite{DBLP:conf/sigir/SachdevaM20,cheng20183ncf}, 
to ensure that each user and item have at least 5 associated reviews. \craig{The resulting Yelp dataset has 47k users, 16k items and 551k reviews; the Amazon dataset has 26k users, 16k items and 285k reviews.} Then, following \cite{DBLP:conf/sigir/SachdevaM20,chen2018neural}, these two datasets are divided into 80\% training, 10\% validation and 10\% \iadh{test} sets in a \craig{time-sensitive} manner. \xwang{In particular, we ensure \iadh{that} the same data split ratio applies to the interactions of each user.} \iadh{Next, we measure} the recommendation effectiveness \wang{by examining if the items interacted with by \iadh{the} users in the test sets are \iadh{actually} chosen for recommendation by the tested models. }
\iadh{Hence}, we \iadh{compute} \craig{the Precision and Recall metrics at different \iadh{standard} rank cutoff positions} (namely, P@1, P@10, and R@10) \iadh{as well as} Mean Average Precision (MAP), \craig{following}~\cite{DBLP:conf/cikm/LiCY17,DBLP:conf/recsys/YangCWT18}, to examine the effectiveness of \iadh{the tested} recommendation approaches. \craig{To test statistical significance, we \iadh{apply a} paired t-test, with significance level to $p< 0.05$}, and \xwang{use the post-hoc Tukey Honest Significant Difference (HSD)~\cite{DBLP:series/irs/Sakai18} at $p < 0.05$ to account for the multiple comparisons with the t-tests.} \iadh{In the following, we describe the experimental} setup of \iadh{both} RPRM and \iadh{the used baselines}.

\begin{table*}[tb]
    \centering
    \caption[Recommendation  performances]{\looseness -1 Recommendation  performances. \craig{Significant differences \iadh{w.r.t.\ `No-Prop'} are indicated by `*' (\craig{according to both} the paired t-test \xw{and the Tukey HSD test}, $p < 0.05$).} \xw{1/2/3 denote a significant difference according to both tests w.r.t.\ to the indicated approach.} \xw{$\uparrow$ \iadh{indicates} \iadh{that} the corresponding approach is significantly outperformed by RPRM on all ranking metrics according to both \craig{tests}.}}
    \resizebox{0.7\linewidth}{!}{%
    \label{tab:property_analysis}
    \resizebox{\columnwidth}{!}{%
    \begin{tabular}{|l||l|l|l|l||l|l|l|l||}
        \hline
        \hline
        \textbf{Dataset} & \multicolumn{4}{c||}{\textbf{Amazon}} & \multicolumn{4}{c||}{\textbf{Yelp}} \\
        \hline
        \hline 
        \textbf{Model} &  \textbf{P@1} & \textbf{P@10} & \textbf{R@10} & \textbf{MAP} &  \textbf{P@1} & \textbf{P@10} & \textbf{R@10} & \textbf{MAP}\\
        \hline
        $\uparrow$BPR-MF & 0.0053* & 0.0034* & 0.0301* & 0.0111* & 0.0101* & 0.0058* &	0.0391* & 0.0145*\\
        $\uparrow$DREAM & 0.0030* & 0.0008* & 0.0062* & 0.0029 * & 0.0083* & 0.0065* & 0.0469* & 0.0155*\\
        $\uparrow$CASER & 0.0093* & 0.0060* & 0.0499* & 0.0239 *& 0.0111* & 0.0083* & 0.0571* & 0.0229*\\
        $\uparrow$1 DeepCoNN & 0.0053*$^{2,3}$& 0.0037*$^{2,3}$ & 0.0343*$^{2,3}$ & 0.0119*$^{2,3}$ & 0.0054*$^{2,3}$ & 0.0025*$^{3}$ & 0.0173*$^{2,3}$ & 0.0072*$^{2,3}$\\
        $\uparrow$2 JRL & 0.0041*$^{1,3}$ & 0.0031*$^{1,3}$ & 0.0310*$^{1,3}$ & 0.0092*$^{1,3}$ & 0.0043*$^{1,3}$ & 0.0021*$^{3}$ & 0.0135*$^{1,3}$ & 0.0061*$^{1,3}$ \\
        $\uparrow$3 NARRE & 0.0175*$^{1,2}$ & 0.0066*$^{1,2}$ & 0.0588*$^{1,2}$ & 0.0279*$^{1,2}$ & 0.0137*$^{1,2}$ & 0.0087*$^{1,2}$ & 0.0605*$^{1,2}$ & 0.0228*$^{1,2}$ \\ 
        \hline
        $\uparrow$No-Prop & 0.0208 & 0.0088 & 0.0805 & 0.0357 & 0.0153 &	0.0099	& 0.0745 & 0.0260\\
        \cdashline{1-9}
        Age & 0.0215* & 0.0089 &	0.0820* & 	0.0372* & 0.0157	& 0.0105*  &	0.0756* & 0.0267*\\
        Length & 0.0214 & 0.0089 & 0.0815* &  0.0364* & 0.0159* & 0.0101 & 0.0726* & 0.0262\\
        Helpful &	0.0218* &	0.0089 & 0.0817* & 0.0365* & 0.0151 & 0.0100 & 0.0719*	& 0.0255\\
        Prob-Helpful & 0.0214 & 0.0093 & 0.0852* & 0.0376*& 0.0152 & 0.0103 & 0.0750	& 0.0264\\
        Rating & 0.0206 & 0.0087 & 0.0795* &	0.0352 & 0.0160*	& 0.0102 & 0.0730* & 0.0264\\
        Polar-Senti & 0.0211 & 0.0086 & 0.0783* & 0.0355 & 0.0155 & 0.0102 & 0.0738 & 0.0262\\
        \cdashline{1-9}
        RPRM & \textbf{0.0223*} &	\textbf{0.0095*} &	\textbf{0.0865*} & \textbf{0.0378*} & \textbf{0.0161*} & \textbf{0.0104} & \textbf{0.0761*} & \textbf{0.0271*}\\
        \hline
    \end{tabular}}
    } 
\end{table*}

\subsection{Model Setting}\label{ssec:model_setting}
\looseness -1 \xiwa{We implement our proposed RPRM model and the NN-based baseline approaches (namely DREAM, CASER, DeepCoNN, JRL and NARRE) using the PyTorch framework~\cite{paszke2019pytorch}.} \iadh{For} the setup of RPRM, in the review processing layer, as introduced in Section~\ref{ssec:RPRM_model}, we use the pre-trained \iadh{BERT} model~\cite{DBLP:conf/naacl/DevlinCLT19} to convert each review into a 768-sized latent vector. \wang{However, since BERT \craig{is limited to encoding a maximum of} to 512 tokens, we limit the maximum review length to be 512 tokens.} \iadh{Next}, in the review property encoding layer, we use the \xiw{Negative Confidence-aware Weakly Supervised (i.e.\ NCWS)} review helpfulness classifier~\cite{wang2020negative} to generate the `Polar\_Helpful' property scores, 
which \craig{estimates} the probability of \iadh{the} reviews being helpful. In particular, we follow~\cite{wang2020negative} \iadh{in training} the NCWS model using reviews from the using `food' and `nightlife' categories of the Yelp Challenge dataset round 12\footnote{We use different Yelp dataset rounds, different categories \& removed reviews that belong to `restaurant' from `food' and `nightlife', to avoid overlaps between the NCWS and RPRM evaluation settings.}  and \wang{on the} Kindle reviews from Amazon\footnote{\craig{Again, we use a different Amazon review category for training NCWS to avoid overlap with the RPRM evaluation.}}. \craig{We use NCWS to predicts the `Polar\_Helpful' property scores of reviews in both the Yelp and Amazon datasets.} 
Similarly, to generate the `Polar\_Senti' review property scores, \wang{we use a CNN-based binary sentiment classifier~\cite{CNN_cls}, which has \iadh{been} shown to have a strong classification accuracy (>95\%)~\cite{wang2019comparison}. We then train it on \craig{50,000  positive and 50,000 negative sentiment reviews} that are sampled from the Yelp Challenge dataset round 12 \xw{to conduct sentiment classification~\cite{wang2019comparison}.}} 
\xiwa{We label the polarity of each review according to the \yaya{user's} posted rating, which \iadh{we label as} positive if the rating $\geq$ 4, and negative if \yaya{the} rating $\leq$ 2.}
This CNN classifier provides each review with its probability of carrying a strong polarised sentiment. \craig{Finally, when training} our proposed RPRM model, we apply \iadh{early-stopping} and use the Adam optimiser~\cite{adam} 
with a $5e^{-4}$ and $1e^{-3}$ learning rates for \iadh{the} Yelp and Amazon datasets, respectively. \xwang{These learning rates are selected after \craig{tuning the model on the validation set, varying the learning rates} between $1e^{-5}$ and $1e^{-3}$.} 

\subsection{Baseline Approaches}
We use \iadh{six baselines: one classical baseline and five state-of-the-art recommendation approaches}: \textbf{(1) BPR-MF}~\cite{DBLP:conf/uai/RendleFGS09} \craig{is a} \iadh{traditional and commonly used} recommendation baseline that uses a pairwise ranking loss function (i.e.\ BPR) to learn the matrix factorised interactions between users and items. \textbf{(2) DREAM}~\cite{yu2016dynamic} encodes the age property of the reviews and \iadh{models} the dynamic representations of \iadh{the} users' preferences with a recurrent neural network. \textbf{(3) CASER}~\cite{tang2018personalized} \craig{is a} \iadh{recent approach} sequentially models \iadh{the} implicit user historical interactions with convolutional neural networks. \iadh{It is} a state-of-the-art recommendation model~\cite{guo2020attentional} that encodes the age property of reviews. \textbf{(4) DeepCoNN}~\cite{zheng2017joint} is a review-based recommendation model that jointly \iadh{models} users and items through a convolutional neural network. \textbf{(5) JRL}~\cite{zhang2017joint} is a heterogeneous recommendation model that encodes various types of information resources including product images, review text and user ratings. In particular, we implement the JRL model by only using the review text. 
\textbf{(6) NARRE}~\cite{chen2018neural} models users and items with two parallel neural \iadh{networks}, \craig{both of which} include a convolutional layer and an attention layer to capture the usefulness of reviews.

\wang{\craig{To ensure a fair} comparison, we also apply early-stopping on all baseline approaches.} In particular, since we use a pre-trained BERT model to convert \iadh{the} reviews into embedding vectors for our proposed RPRM model, we also \iadh{extend} the DeepCoNN and NARRE baselines \iadh{by} using the BERT-encoded review embedding vectors. \xwang{In particular, for DeepCoNN, we concatenate all reviews given by/to a single user/item and form a user/item review document.} \craig{Similar to RPRM, for both approaches, we limit the maximum length of \iadh{the} user/item document in \iadh{the} DeepCoNN model, and the maximum tokens of each review in NARRE, to be 512 tokens.}
\wang{We then fine tune every baseline model with learning rates in $[1e^{-3}, 1e^{-4}, 1e^{-5}]$ and we compare our approaches with \craig{the settings that exhibited the} best performances \xwang{on the validation set.}}  \xiwa{Furthermore, we incrementally evaluate the various components of our proposed RPRM model. First}, we capture the effectiveness of using \iadh{the} review properties in a review-based recommendation model, we remove the review property encoding layer in RPRM, \iadh{denoted as} `No-Prop', to examine its effectiveness on \iadh{our} datasets. \iadh{Next}, we also examine the effectiveness of using each single review property from the included properties in Section~\ref{ssec:RPRM_model}. Therefore, we apply each single review property in the review property \xwang{attention} layer of RPRM to \wang{evaluate} \wang{their} \wang{effectiveness in identifying the} usefulness of reviews. \iadh{We denote the resulting} recommendation models with the name of the corresponding review properties (e.g.\ `Age' for using the Age property). \xiwa{\yaya{Next}, we examine the effectiveness of our proposed RPRM's learning \xiwa{schemes}, namely \yaya{the} two loss functions ($PropLoss_{uu}$ and $PropLoss_{ui}$) and \iadh{the} negative sampling strategy $PropSample$, by comparing \yaya{their performances} \yaya{with those of} the commonly used BPR loss function and \yaya{with} uniform sampling, \yaya{respectively}.} 


\section{Results and Analysis}\label{sec:result_analysis}
\looseness -1 \iadh{We} present and analyse the results of our experiments to answer the research questions in Section~\ref{sec:experimental_setup}. \iadh{Our experiments focus on} investigating the performance of \iadh{RPRM }\iadh{as well as the effectiveness of our} proposed loss functions and negative sampling strategies \iadh{in comparison to the \xwang{six} strong baselines from the literature}.

\subsection{RQ1: Review Property-based Model Evaluation}
\iadh{To answer RQ1, we first examine the performances \iadh{of} the six \iadh{baselines} and compare them to the performance of our proposed RPRM model and its variants. In particular, we integrate each review property separately, before \craig{combining} all of them together in the full RPRM model. The results on the two used datasets are presented in Table~\ref{tab:property_analysis}}.
First, \iadh{we compare the performance} of the RPRM model without using \iadh{the} review property \iadh{encoding layer} (namely No-Prop) to the baseline approaches from Table~\ref{tab:property_analysis}. We observe that No-Prop significantly outperforms all baseline approaches, \xiwa{including the state-of-the-art recommendation approaches (namely CASER, DeepCoNN and NARRE),} \xw{according to both \iadh{the} paired t-test and the Tukey HSD test} \iadh{regardless of whether they use any} review information. In particular, we focus on DeepCoNN, JRL and NARRE that \craig{make use of} review information. Both DeepCoNN and JRL \craig{exhibit} weak recommendation performances on \iadh{the} two \iadh{used} datasets with low precision, recall and MAP scores, which are lower than the \iadh{traditional} BPR-MF approach. \xiwa{The BPR-MF approach is a strong baseline and \yaya{was shown recently to outperform} various state-of-the-art recommendation approaches from the literature~\cite{DBLP:conf/recsys/RendleKZA20}.}
Among these three baselines, NARRE \xw{significantly} outperforms both \iadh{the} DeepCoNN and JRL approaches \xw{according to both \iadh{the} paired t-test and the Tukey HSD test} on \iadh{the} two datasets with higher evaluation scores. 
However, NARRE is \xwang{significantly} \iadh{outperformed} by \iadh{our} No-Prop \iadh{variant} (\xw{according to both \iadh{the} paired t-test and the Tukey HSD test}), \craigw{despite No-Prop having} a simpler structure than NARRE. The effectiveness of \iadh{this} simple review-based recommendation \iadh{approach} \iadh{is consistent} with the conclusions in~\cite{DBLP:conf/sigir/SachdevaM20}. Moreover, by comparing No-Prop and DeepCoNN, \iadh{we note that} \iadh{the} only architecture difference between these two models is that No-Prop integrates the identification embedding vectors of users and items. The \iadh{observed} significantly \craig{enhanced performances} of No-Prop \iadh{over \xwang{DeepCoNN} \xw{on all \iadh{used} metrics (paired t-test and Tukey HSD test}) \craig{demonstrate}} the benefits of using such embedding vectors to model \iadh{the} users' \iadh{preferences} and items' \craig{attributes}. \craigw{In summary, we find that} \xiw{No-Prop significantly outperforms all baseline approaches. \yaya{In particular}, the use of the embedding vectors, which model the users' preferences and items' attributes, \yaya{explains the} superior 
performances of both \craig{the} No-Prop and NARRE models \xiwa{\yaya{in comparison} to other baseline approaches}.}

\begin{table*}[tb]
    \centering
    \caption{\iadh{Impact of the} \xiwa{model's learning schemes} on \craig{RPRM}. \xw{Statistically significant differences with respect to `RPRM-basic' are indicated by `*' (according to both the paired t-test and the Tukey HSD test, $p < 0.05$).}}
    \resizebox{0.6\linewidth}{!}{%
    \label{tab:model_learning_impacts}
    \begin{tabular}{|m{2.4cm}||m{0.75cm}|m{0.75cm}|m{0.75cm}|m{0.75cm}||m{0.75cm}|m{0.75cm}|m{0.75cm}|m{0.75cm}||}
        \hline
        \hline
        \textbf{Dataset} & \multicolumn{4}{c||}{\textbf{Amazon}} & \multicolumn{4}{c||}{\textbf{Yelp}} \\
        \hline
        \hline 
        \textbf{Model} &  \textbf{P@1} & \textbf{P@10} & \textbf{R@10} & \textbf{MAP} &  \textbf{P@1} & \textbf{P@10} & \textbf{R@10} & \textbf{MAP}\\
        \hline
        RPRM-basic & 0.0223 & 0.0095 & 0.0865 & 0.0378 & 0.0161 & 0.0104 & 0.0761 & 0.0271 \\
        \cdashline{1-9}
        $PropLoss_{uu}$-KL &	0.0217 & 0.0094 & 0.0867 & 0.0381 & 0.0163 & 0.0106 & 0.0772* & 0.0281*\\
        $PropLoss_{uu}$-Cos & 0.0211* & 0.0093 & 0.0853* & 0.0369* & 0.0154*& 0.0105 & 0.0764	& 0.0271\\
        $PropLoss_{ui}$-KL &	\textbf{0.0226} & \textbf{0.0097}	& \textbf{0.0894}* & \textbf{0.0385}* & \textbf{0.0175*} &	\textbf{0.0107} & \textbf{0.0788*}	& \textbf{0.0288*}\\
        $PropLoss_{ui}$-Cos & 0.0211* & 0.0093 & 0.0859* & 0.0382 & 0.0165 & 0.0105 & 0.0772* & 0.0278*\\
        \cdashline{1-9}
        $PropSample$-KL &	0.0225 & 0.0093 & 0.0863 & 0.0385* & 0.0163 &  0.0102 & 0.0738* & 0.0268\\
        $PropSample$-Cos & 0.0220 & 0.0093 & 0.0855 & 0.0376& 0.0166 & 0.0105 & 0.0767* & 0.0276\\
        \cdashline{1-9}
        $PropLoss_{ui}$-KL +$PropSample$-KL&	0.0210* & 0.0095 & 0.0871* & 0.0373 & 0.0162 & 0.0100 & 0.0740* & 0.0266\\
        \cdashline{1-9}
        $PropLoss_{ui}$-KL +$PropSample$-Cos & 0.0211* & 0.0094 & 0.0863	& 0.0372* & 0.0159 & 0.0105 & 0.0770* & 0.0273\\
        \hline
    \end{tabular}%
    }
\end{table*}

\iadh{Next}, we evaluate the effectiveness of integrating different review properties to \iadh{the basic} No-Prop \iadh{approach} to model the usefulness of reviews. The results from Table~\ref{tab:property_analysis} show that \wang{in general}, the review properties can \xwang{significantly} improve \iadh{the} performances of No-Prop \iadh{on both used datasets} \xw{according to both the paired t-test and the Tukey HSD test}.
\iadh{In particular, we observe that the} `Age' and `Prob\_Helpful' review properties are the two most effective properties \iadh{among the six review properties we tested} in capturing the usefulness of reviews and improving the recommendation \craig{effectiveness} of No-Prop. \iadh{The other} review property-based approaches show different performances on \iadh{the} two datasets. For example, by the `Helpful' property \iadh{enhances} the performances of No-Prop on the Amazon dataset but \iadh{decreases} its performances on the Yelp dataset. Moreover, the \craig{`Rating'} property \iadh{improves} the recommendation \iadh{performances} of No-Prop on the Yelp dataset but not on the Amazon dataset. Therefore, these results \iadh{suggest} that it is \iadh{more} effective to \iadh{selectively apply the right} review properties in \iadh{the} recommendation model to \iadh{assess} the usefulness of the reviews and \iadh{leverage them in the made} recommendations\wang{, which is one of \iadh{the main underlying ideas} of proposing \iadh{the} RPRM model.} In particular, \iadh{these} results indicate the necessity of understanding the importance of different review properties on different datasets or recommendation applications. Therefore, next, we evaluate the performance of our proposed RPRM model, which integrates all six review properties and \iadh{appropriately scores} (or weights) the importance of different reviews' properties. The \iadh{observed} results \iadh{for} RPRM from Table~\ref{tab:property_analysis} show that the RPRM \iadh{model} provides the best recommendation \craig{effectiveness} on \iadh{the} two \iadh{used} datasets. \iadh{Moreover, the} observed \iadh{performances}  significantly outperform \iadh{both} No-Prop \wang{and \iadh{all the} baseline approaches\xiwa{, including the \yaya{existing} state-of-the-art recommendation \yaya{models} (namely NARRE, CASER and DeepCoNN),})} \xw{according to both \iadh{the} paired t-test and the Tukey HSD test}.
These results \iadh{demonstrate} the benefits of using all review properties and weighting their importance \iadh{for} capturing \iadh{the} useful reviews \iadh{and their leverage in recommendation}.

Therefore, \iadh{in answering} RQ1, we conclude that different review properties show distinct effectiveness \iadh{levels in enhancing} the performance of a review-based recommendation model. \iadh{Among the six used} review properties, the `Age' and `Prob\_Helpful' \iadh{properties} are the most effective, \iadh{and} consistently enhance the effectiveness of the No-Prop recommendation model. Furthermore, by integrating all six review properties and weighting their importance in \iadh{the full} RPRM \iadh{model}, \iadh{we observe that} RPRM \iadh{achieves the} best performance \iadh{among all tested approaches} on \iadh{both used} datasets. \xw{\iadh{Our results} also \iadh{validate} \iadh{our} hypothesis in Section~\ref{sssec:attention_layer}, \iadh{namely} that weighting the importance of review properties with a dot-product attention mechanism can enhance the recommendation performances.} 


\begin{figure*}[tb]
\def\HMfigurewidth{53mm}
    \centering
\begin{subfigure}{\HMfigurewidth}
\centering
    \includegraphics[width=\HMfigurewidth]{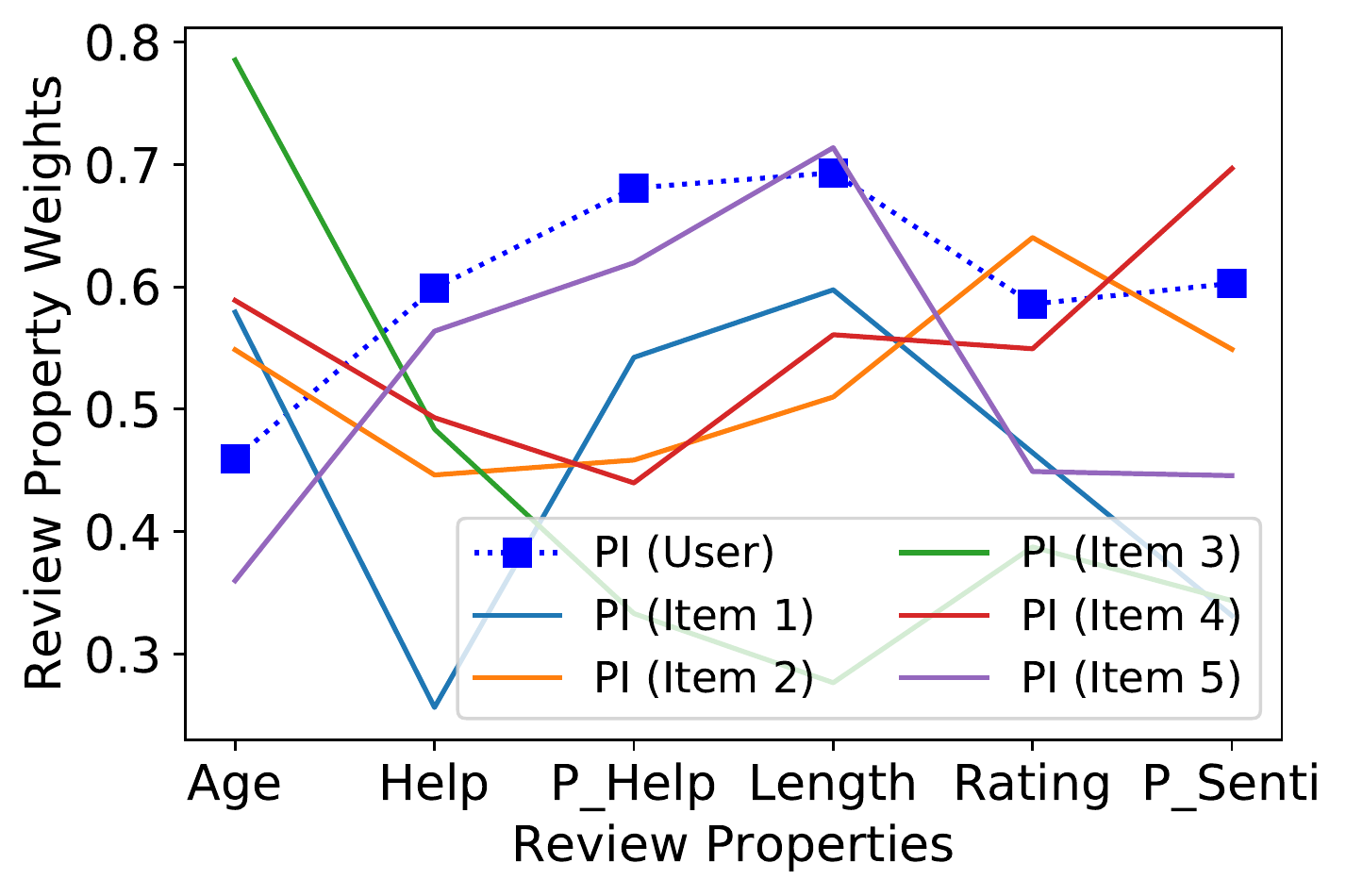}
    \caption{User A (few interactions).}
    \label{fig:user_a}
\end{subfigure}
\begin{subfigure}{\HMfigurewidth}
    \centering
    \includegraphics[width=\HMfigurewidth]{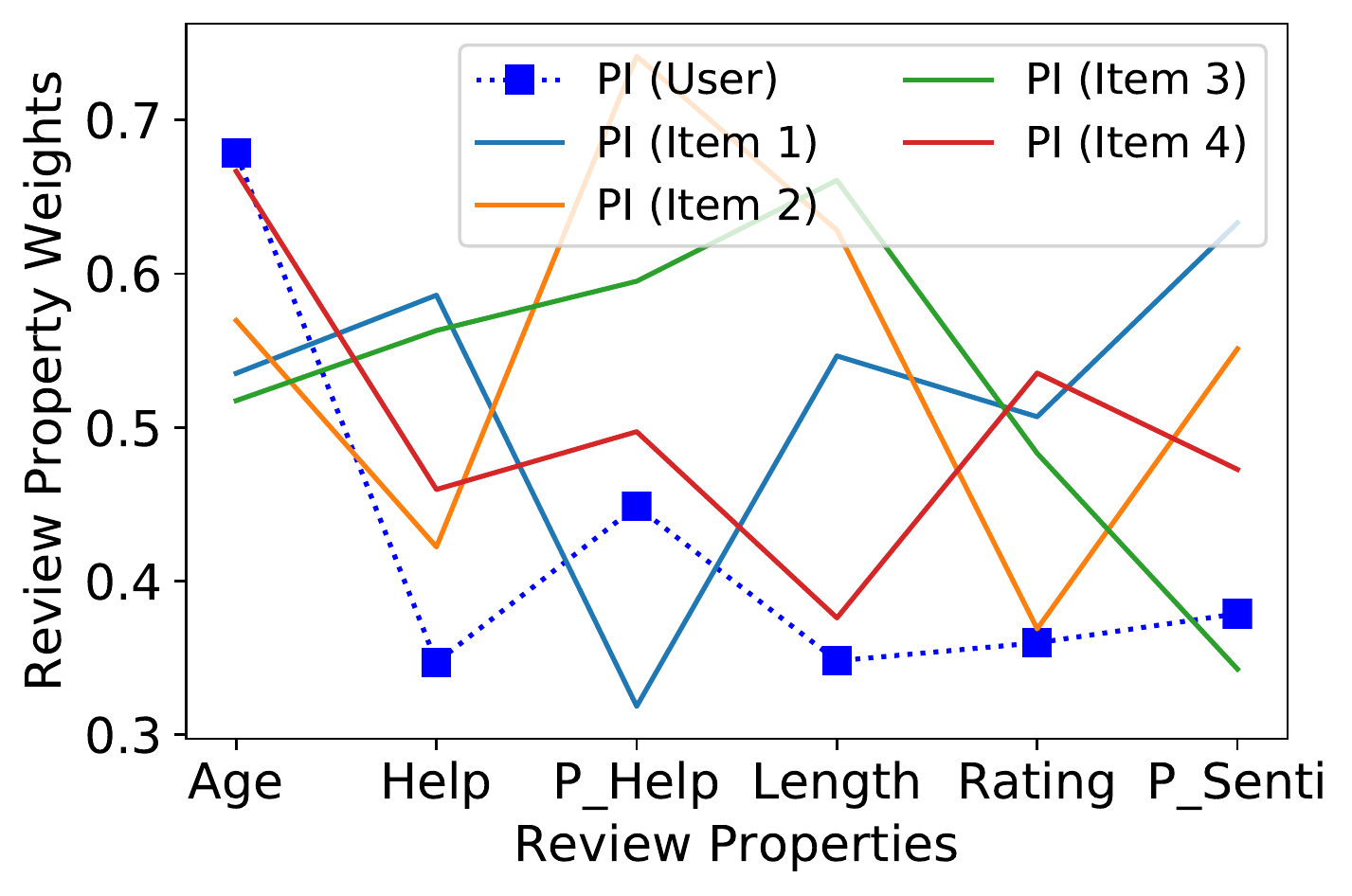}
    \caption{User B (few interactions).}
    \label{fig:user_b}
\end{subfigure}
\begin{subfigure}{\HMfigurewidth}
    \centering
    \includegraphics[width=\HMfigurewidth]{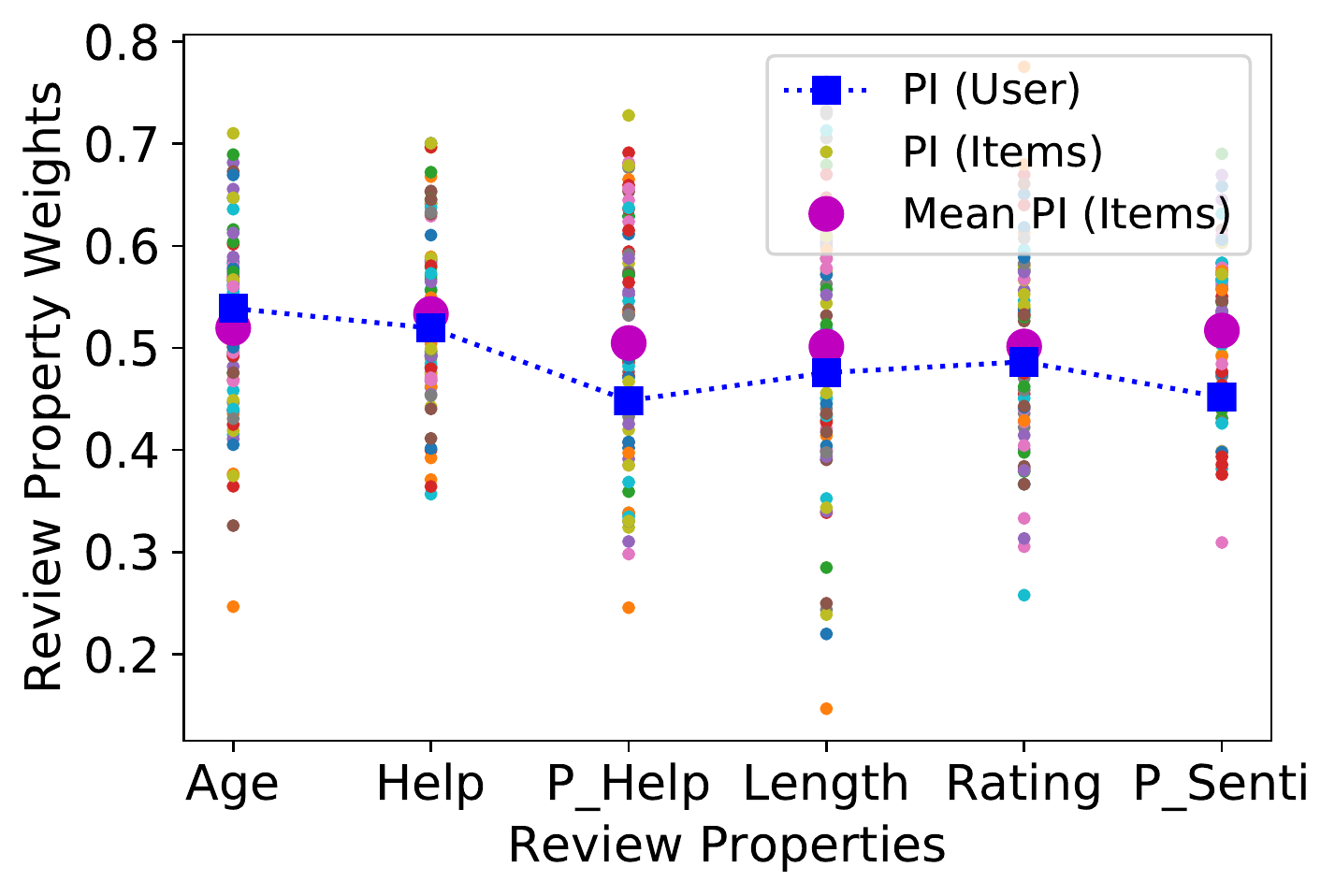}
    \caption{User C (many interactions).}
    \label{fig:user_c}
\end{subfigure}
\caption{The \xw{\xiwa{properties'} importance scores of reviews} \iadh{for} \xiwa{randomly selected} users and their interacted items. `Help', `P\_Help', `P\_Senti' are the abbreviations of `Helpful', `Prob\_Helpful' \& `Polar\_Senti', resp. \xw{`PI' refers to the \xiwa{Properties'} Importance scores.}} 
\label{fig:user_examples}
\end{figure*}

\subsection{RQ2: Effectiveness of \iadh{the} Proposed Loss Functions} \label{SS:LossF} 
\looseness -1 \iadh{To answer} RQ2, we examine the impact of using our proposed loss functions (namely $PropLoss_{uu}$ and $PropLoss_{ui}$) using two different similarity approaches (\craig{namely the} KL divergence and \craig{Cosine} similarity). We also compare \iadh{the} RPRM model that uses our proposed loss functions \iadh{with the same model using a standard} ranking scheme, \iadh{namely the} 
\craig{BPR} loss function and \iadh{a} uniform sampling strategy \iadh{for} generating negative items (i.e.\ RPRM-basic). \iadh{By} exploring different combinations, we have four \iadh{possible} model learning setups, \wang{i.e.\ $PropLoss_{uu}$ with KL or \craig{Cosine} and $PropLoss_{ui}$ with KL or \craig{Cosine}}. \iadh{Table~\ref{tab:model_learning_impacts} presents the} \iadh{obtained} experimental results on \iadh{the} two Amazon and Yelp datasets \iadh{for} these four model learning setups. First, from Table~\ref{tab:model_learning_impacts}, we observe that our proposed two loss functions can consistently improve the performance of the basic RPRM with \iadh{the exception of the} $PropLoss_{uu}$-Cos \iadh{model setup} on the Amazon dataset. In particular, by comparing the evaluation \iadh{performances} \iadh{of} the PropLoss-based approaches \iadh{with that of} the basic RPRM, \iadh{we observe that} $PropLoss_{ui}$-KL \craig{improves upon} the recommendation performances of RPRM-basic with significantly higher MAP scores \xw{according to both \iadh{the} paired t-test and the Tukey HSD test} on \iadh{the} two \iadh{used} datasets: 0.03784 $\rightarrow$ 0.03857 on \iadh{the} Amazon dataset and 0.02713 $\rightarrow$ 0.02880 on \iadh{the} Yelp dataset, \iadh{which is} significant according to \xw{both the paired t-test and the Tukey HSD test} at \xwang{$p<0.05$}. 

\looseness -1 \iadh{Next}, we compare the impact of \iadh{integrating} \iadh{the} two \iadh{proposed} loss functions \iadh{in turn} \iadh{into} RPRM.
\iadh{From} the results \iadh{in} Table~\ref{tab:model_learning_impacts}, \craig{we observe that} \iadh{the} $PropLoss_{ui}$-based approaches \iadh{outperform the} $PropLoss_{uu}$-based approaches on \iadh{both} datasets. \iadh{This observation suggests} that, \iadh{in terms of setting the importance of the used review properties}, it is more effective to amplify the disagreement \iadh{between users and the negatively sampled items} than \iadh{that} between the users' interacted items and the negatively sampled items. \xw{\iadh{Our results} also \iadh{demonstrate} \iadh{that} leveraging the users' adoption of information framework \iadh{is a promising approach}}.
\iadh{Finally}, by examining the effectiveness of the two similarity measurement approaches, we observe that both \iadh{the} KL and \craig{Cosine} similarity-based approaches can outperform the RPRM-basic \iadh{model} when \iadh{applied with} the $PropLoss_{ui}$ approaches. However, KL is consistently more effective on \iadh{both} datasets \iadh{in comparison to} the \craig{Cosine} similarity method \xwang{and significantly better than RPRM-basic} \xw{according to both the paired t-test and the Tukey HSD test}.
\iadh{This result overall demonstrates} the effectiveness of modelling the divergence similarity between the weighting vectors of \iadh{the} review properties \iadh{on our used datasets}.

\looseness -1 \craig{After} analysing the results from Table~\ref{tab:model_learning_impacts}, we now answer RQ2: our proposed loss functions $PropLoss_{uu}$ and $PropLoss_{ui}$ can both improve the performances of RPRM-basic. Moreover, $PropLoss_{ui}$ shows \iadh{a} \xwang{higher} effectiveness than $PropLoss_{uu}$ \iadh{in enhancing the recommendation performance}. \iadh{We conclude that} the divergence between \iadh{the weighted} vectors of \iadh{the} review properties \iadh{using} the KL divergence \wang{measure}, $PropLoss_{ui}$, can enhance the RPRM-basic \iadh{model} \craig{and} \iadh{gives} the best \iadh{overall} recommendation \iadh{performances}.

\vspace{\baselineskip}
\subsection{RQ3: Effectiveness of \iadh{the} Proposed Negative Sampling Strategy}
\looseness -1 \iadh{We now examine the effectiveness} of our proposed negative sampling strategy (namely $PropSample$), \iadh{so as to} answer RQ3.  \craig{From} Table~\ref{tab:model_learning_impacts}, we observe that $PropSample$ does not consistently outperform the RPRM-basic \iadh{model} \iadh{when using} the same similarity approach. \iadh{In particular,} the $PropSample$ \iadh{model} with the KL divergence can improve the recommendation performance of RPRM-basic on \iadh{the} Amazon dataset but not on \iadh{the} Yelp dataset. On the other hand, the $PropSample$ \iadh{model} that uses the \craig{Cosine} similarity can improve the recommendation performance of RPRM-basic on \iadh{the} Yelp dataset but not on \iadh{the} Amazon dataset. 
\iadh{Next, we investigate combining the $PropSample$ negative sampling strategy with the best performing loss function, 
namely $PropLoss_{ui}$-KL (see the last two rows of Table~\ref{tab:model_learning_impacts})}. \iadh{We observe that the combination of both $PropSample$ and $PropLoss_{ui}$-KL with RPRM-basic does not lead to \iadh{a} better performance than \iadh{when} solely using $PropLoss_{ui}$-KL.}
 \iadh{These} results \iadh{might be} caused by the \wang{fact} that both $PropLoss_{ui}$ and $PropSample$ \wang{similarly} capture the importance of \iadh{the} review properties between \iadh{the} users' interacted items and \iadh{the} unseen items. 
Furthermore, 
$PropSample$ only considers the agreement \iadh{between \iadh{the} users' interacted \iadh{(positive)} and \iadh{the} unseen \iadh{(negative)}} items on the \iadh{important reviews' properties}, which might not be sufficient to sample \xwang{useful} negative items. We \craig{leave the modelling of \wang{further} additional information in the $PropSample$ \wang{approach}} \wang{(e.g. the agreement on the \iadh{important reviews' properties} between the users and their interacted \iadh{(positive)} \iadh{and/or} unseen \iadh{(negative)} items)} \iadh{to future work}.

\looseness -1 Therefore, for RQ3, we conclude that $PropSample$ can 
enhance RPRM
\iadh{if an adequate similarity measure is applied on each used dataset.} 
\wang{In \iadh{addition}, by comparing the performances \iadh{of} the $Prop$ -$Sample$ \iadh{and} $PropLoss$ approaches, \iadh{our results showed that the $PropLoss$ loss function has more \craig{impact} on the recommendation effectiveness than the negative sampling strategy, suggesting that it is more important to capture the reviews' properties importance between the users and their interacted or unseen items}.} 



 \section{Users' Property Preferences}
\looseness -1 \craig{The users' adoption of information is one of the main arguments \craig{underlying} \iadh{our proposed RPRM model}. In Section~\ref{sec:result_analysis}, we \craig{showed} \iadh{the} effectiveness of \iadh{modelling the agreement between the users and items} 
\craig{in terms of the reviews' properties.} 
Therefore, in this section, we use \xw{three} \craig{\xiwa{randomly selected} users} to \iadh{illustrate} the users' preferences on different \iadh{review} properties \wang{and the agreement on the importance of review properties between the users and their interacted items.} }

\looseness -1 \craig{To this end, Figure~\ref{fig:user_examples}(a)-(c) plots the learned RPRM \xw{property importance} scores \iadh{for the} review properties \iadh{of} three \xiwa{randomly selected} users, say A, B \& C, as well as their interacted items. The users' property importance preferences are shown using a blue dashed line with square markers; their interacted items in the \iadh{test} set \iadh{are also shown} (solid lines in Figure~\ref{fig:user_a} and~\ref{fig:user_b} and dots in different colours in Figure~\ref{fig:user_c}) from the Amazon dataset.}
%
%
%
\craig{In particular, we selected users A \& B from the Yelp \iadh{dataset} as \iadh{example} users that have few interactions with items, to \wang{illustrate}  the \xw{importance} scores on \iadh{the} review properties between the target users and their interacted items. Indeed, we selected users with few interactions so as to be able to visually plot all these items in a figure.} 
From Figure~\ref{fig:user_a}, we observe that user A shows stronger preferences \iadh{for} the `Length' property \wang{(i.e.\ A prefers longer reviews)} and \craig{that} the `Length' \wang{property is \xwang{also highly weighted when determining the usefulness \iadh{of} reviews associated to that user's interacted items.}} 
On the other hand, \iadh{for} user~B \craig{(Figure~\ref{fig:user_b})}, the `Age' property is an important review property \wang{(i.e.\ B prefers recent reviews)} to capture \iadh{the} review usefulness, \craig{which is similar to the high weights on `Age' for the interacted items.} 

\looseness -1 Next, since \craig{users} A and  B have few item interactions, \iadh{they might not accurately reflect the \craig{behaviour of the general user population} in} using different reviews. Therefore, we plot the learned \xw{importance scores} of \iadh{the} review properties \wang{for} \craig{a third} user, C, \craig{who} \iadh{has} \craig{interacted with} a higher \iadh{number} of items. We also \wang{plot the \craig{mean} importance scores of} \iadh{the} review properties of his/her interacted items in Figure~\ref{fig:user_c}. From Figure~\ref{fig:user_c}, we observe that the importance scores on \iadh{the} review properties of user C is close to the \craig{mean} importance scores on \iadh{the} review properties of his/her interacted items, especially on the two most important review properties (`Age' and `Helpful'). \wang{\craig{This tells us that} the 'Age' and 'Helpful' properties are important properties to observe the usefulness of reviews for both user C and his/her interacted items.} 


\craig{The above figures provide further evidence} that users and their interacted items \xwang{agree on the important review properties}. 
\iadh{We envisage} that an online platform could leverage these weighting scores to customise the review presentation to different users according to their \iadh{preferences for} different review properties. For example, it is better to \xwang{present} recent reviews to user B than to user A, \iadh{while} user A would prefer to \iadh{see longer} reviews \iadh{so as to obtain more information about the items' features}. \wang{Our proposed RPRM model can learn the importance of the review properties to \iadh{identify} useful reviews 
and \iadh{enables} making review presentation decisions.}



\section{Conclusions}

\iadh{We} \iadh{proposed} the review-based RPRM model, which \iadh{leverages} the \iadh{importance} of different review properties in capturing the usefulness of reviews \iadh{thereby enhancing the recommendation performance}. \iadh{Inspired by the users' adoption of information \xwang{framework}}~\cite{sussman2003informational}, we \iadh{proposed} two new loss functions and a negative sampling strategy \iadh{that account for} 
the usefulness of \iadh{the} review properties. RPRM \iadh{consistently} outperformed six \iadh{strong} recommendation approaches \iadh{across the two used} datasets. Moreover, \iadh{we have shown that} both of our proposed loss functions and negative sampling strategy can further improve the recommendation \iadh{performances} of RPRM. \iadh{These results demonstrated} \xwang{the \iadh{advantages} of leveraging the agreement on the \iadh{review properties' importance} between users and items.} 
\iadh{Through a qualitative analysis, we have also illustrated} the recommendation \iadh{added-value} of RPRM by examining the usefulness of several review properties \iadh{for a sample of} users and their interacted items. This \iadh{analysis has exemplified the promise of RPRM in guiding online review platforms in customising} the presentation of reviews and \iadh{deploying} more effective recommendation systems.

\balance
\bibliographystyle{ACM-Reference-Format}
\bibliography{sample-base}
\end{document}